# Interface-induced Polarization in SrTiO$_3$-LaCrO$_3$ Superlattices

*Ryan B. Comes,$^{1}$* Steven R. Spurgeon,$^{1}$ Steve M. Heald,$^{2}$ Despoina M. Kepaptsoglou,$^{3}$ Lewys Jones,$^{4}$ Phuong Vu Ong,$^{1}$ Mark E. Bowden,$^{5}$ Quentin M. Ramasse,$^{3}$ Peter V. Sushko,$^{1}$ and Scott A. Chambers$^{1}$**

Drs. R.B. Comes, S.R. Spurgeon, P.V. Ong, P.V. Sushko, S.A. Chambers
Physical and Computational Sciences Directorate, Pacific Northwest National Laboratory
Richland, WA 99352, USA
E-mail: ryan.comes@pnnl.gov; sa.chambers@pnnl.gov

Dr. S.M. Heald
Advanced Photon Source, Argonne National Laboratory,
Argonne, Illinois 60439, USA

Drs. D.M. Kepaptsoglou, Q.M. Ramasse
SuperSTEM, SciTech Daresbury Campus,
Daresbury, WA44AD, United Kingdom

Dr. L. Jones
Department of Materials, University of Oxford,
Oxford OX13PH, UK

Dr. M.E. Bowden
Environmental Molecular Sciences Laboratory, Pacific Northwest National Laboratory
Richland, WA 99352, USA

Epitaxial interfaces and superlattices comprised of polar and non-polar perovskite oxides have generated considerable interest because they possess a range of desirable properties for functional devices. In this work, emergent polarization in superlattices of SrTiO$_3$ (STO) and LaCrO$_3$ (LCO) is demonstrated. By controlling the interfaces between polar LCO and non-polar STO, polarization is induced throughout the STO layers of the superlattice. Using x-ray absorption near-edge spectroscopy and aberration-corrected scanning transmission electron microscopy displacements of the Ti cations off-center within TiO$_6$ octahedra along the superlattice growth direction are measured. This distortion gives rise to built-in potential gradients within the STO and LCO layers, as measured by *in situ* x-ray photoelectron spectroscopy. Density functional theory models explain the mechanisms underlying this behavior, revealing the existence of both an intrinsic polar distortion and a built-in electric field, which are due to alternately positively and negatively charged interfaces in the superlattice. This study paves the way for controllable polarization for carrier separation in multilayer materials and highlights the crucial role that interface structure plays in governing such behavior.

**1. Introduction**

Complex oxide superlattices comprised of dissimilar materials exhibit a wide range of novel structural, electronic and magnetic properties due to the high density of interfaces present in such thin films. The large number of interfaces in epitaxial superlattices can give rise to emergent properties within the interior of the multilayer that may not be present or measureable at a single interface. These include distortions of the oxygen octahedral sub-lattice due to different octahedral tilts between the two materials,[1–3] and charge transfer due to band alignment across the interface of isovalent complex oxides.[4,5] In particular, superlattices combining both ferroelectric and a non-ferroelectric oxides have generated a great deal of interest due to induced ferroelectric polarization in the non-ferroelectric layer of the superlattice.[6] These systems include $PbTiO_3$-$SrTiO_3$[7,8] and $BaTiO_3$-$SrTiO_3$,[9,10] where a polarization was induced in the $SrTiO_3$ (STO) layer. Others have explored $PbTiO_3$-$SrRuO_3$ superlattices and shown that metallic $SrRuO_3$ behaves as an insulator along the superlattice direction due to the $PbTiO_3$ polarization.[11] However, to date there has been no work exploring polarization induced at interfaces between two non-ferroelectric oxides. In particular, there is a rather poor understanding of local ordering in such systems and of how surface termination and band alignment affect the overall superlattice behavior.

In contrast to isovalent superlattices, limited work has been carried out on superlattices comprised of aliovalent *A*- and *B*-site cations. Such a lattice consists of consecutive layers of a non-polar perovskite with chemical formula $A^{2+}B^{4+}O_3$ and a polar perovskite with the formula $A'^{3+}B'^{3+}O_3$. STO-$LaMnO_3$ superlattices have been studied to examine charge leakage across the interface,[12] as well as the novel optical states that emerge at the interface.[13] Recent work examining LaCoO$_3$-STO superlattices focused on variation of octahedral tilt angles across the superlattice, but did not indicate any polarization in STO.[3] LaAlO$_3$ (LAO) and STO superlattice quantum wells have shown unique optical properties due to quantum confinement.[14] LAO-STO superlattices have also been explored using x-ray scattering to examine the role of interfacial oxygen vacancies on the electronic structure of STO at the interface.[15] A switchable polarization at a single LAO-STO interface grown on (La,Sr)(Al,Ta)O$_6$ (LSAT) has also been observed using piezoresponse force microscopy and attributed to oxygen vacancy migration.[16,17] STO-LaCrO$_3$ (LCO) superlattices with a periodicity of 1 unit cell (u.c.) of STO and 1 u.c. of LCO along the (001) direction are predicted to exhibit unique optical properties that would occur with a high density of Ti-O-Cr bonds along the growth direction.[18] Additionally, the previously observed potential gradient within LCO films grown on STO[19] offers exciting possibilities in LCO-STO superlattices as it may be possible to achieve a potential gradient within the confined STO layer in a superlattice as well. In contrast, the absence of a measureable potential gradient at the LAO-STO interface has been the subject of much controversy in understanding phenomena that occur in that system.[20,21]

Here, we present a combined experimental and theoretical investigation of the properties of STO-LCO superlattices. We demonstrate that it is possible to induce a ferroelectric distortion throughout the STO layer by engineering the growth process to produce alternating positively and negatively charged interfaces. This configuration is shown schematically and in a high-angle annular dark-field scanning transmission electron (STEM-HAADF) micrograph in **Figure 1**(a) along with an out-of-plane x-ray diffraction scan for a 6 unit cell (u.c.) STO-3 u.c. LCO (STO$_6$-LCO$_3$) superlattice with a 3 u.c. STO capping layer grown on LSAT (001) in Figure 1(b). The x-ray diffraction pattern shows clear superlattice peaks, indicating a uniformly repeating structure throughout the film. Superlattices with 8 u.c. of STO-4 u.c. of LCO and 4 u.c. of STO-2 u.c. of LCO were also investigated, but the majority of our discussion and characterization will focus on the STO$_6$-LCO$_3$ superlattice. By engineering the interfaces, we show that it is possible to produce a built-in electric field within the STO layer, resulting in a polarization that we observe both experimentally and via simulations. Our results provide exciting new insights into the local mechanisms governing such behavior and open the door to the engineering of emergent polarization in heterostructures, which may be useful for charge separation in photochemical and photovoltaic applications.

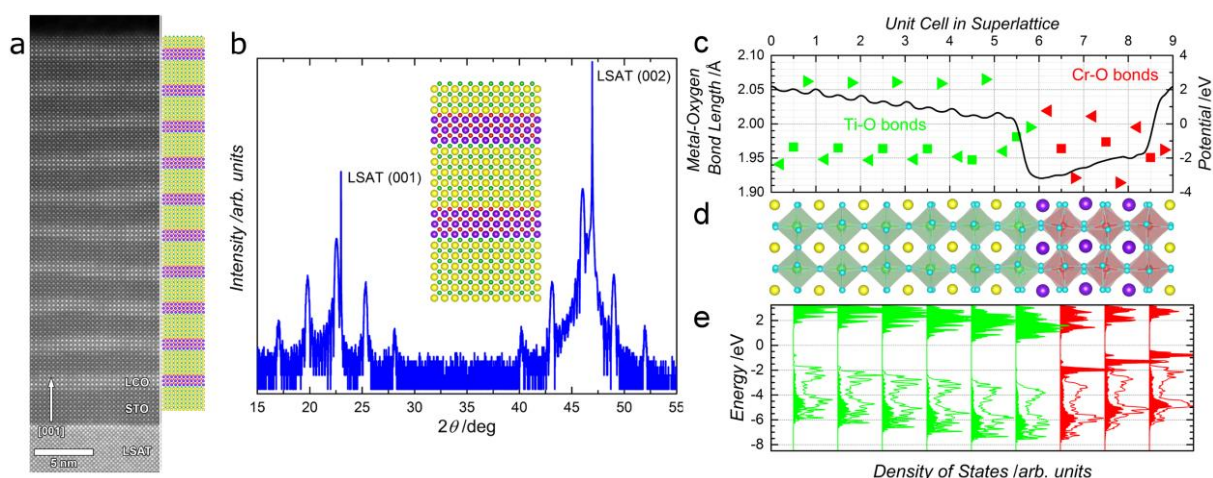

**Figure 1.** a) Representative high-angle annular dark field (STEM-HAADF) micrograph and model of 6 u.c. SrTiO$_3$- 3 u.c. LaCrO$_3$ superlattice with STO cap. Yellow: Sr; Green: Ti; Purple: La; Red: Cr. b) Out-of-plane x-ray diffraction of the STO$_6$-LCO$_3$ superlattice with LSAT substrate peaks noted; inset shows larger version of superlattice. (c-e) Density functional theory model of a 6 u.c. SrTiO$_3$-3 u.c. LaCrO$_3$ superlattice showing (c) the electrostatic potential (black line) and the transition metal-oxygen bond lengths for axial (triangles, pointing towards apical bond) and in-plane (squares) bonds for Ti and Cr; (d) structural model of superlattice showing octahedral behavior; and (e) layer projected density of states on $B$O$_2$ planes, with states associated with transition metals shaded.

## 2. Results
### 2.1. *Ab Initio* Modeling

A density functional theory (DFT) model (described in the Experimental Section) of the idealized STO$_6$-LCO$_3$ superlattice was constructed to examine the electronic and structural behavior in the STO and LCO layers of the superlattice. The structural model of the STO$_6$-LCO$_3$ superlattice, density of states projected on the atomic orbitals of the $B$O$_2$ layers ($B$ = Ti and Cr), metal-oxygen bond lengths, and electrostatic potential are shown in Figure 1. The potential (**Figure 1**(c)) is averaged over each LCO unit cell in the superlattice, but shows slight artificial oscillations within the STO layer due to the differences in out-of-plane lattice parameter between STO and LCO. The apical (triangles) and in-plane (squares) bond lengths for each $BO_2$ layer of the superlattice are also shown in **Figure 1**(c). The relaxed structure is seen in **Figure 1**(d) and is aligned to the modeling results in the graphs above and below. Within the periodic boundary conditions we used, the right-most SrO layer (unit cell 9) is equivalent to the left-most (unit cell 0) layer. Throughout this work we will refer to the interfaces at unit cell 9 between the $CrO_2$ layer and the SrO layer as being negatively charged due to the net -$e$ charge on the $Cr^{3+}(O^{2-})_2$ layer and the neutral $Sr^{2+}O^{2-}$ layer. Likewise, we will refer to the $La^{3+}O^{2-}$ -$Ti^{4+}(O^{2-})_2$ interface as being positively charged due to the net +$e$ charge at this interface. **Figure 1**(e) shows the projected density of states that matches the behavior of the potential. The model indicates built-in electric fields of opposite signs across both the STO and LCO. The built-in potential gradient in the LCO layer is consistent with what has previously been observed experimentally and predicted theoretically in LCO films grown on STO substrates.[19] Due to the confined nature of the superlattice, the STO layer is perturbed out of its equilibrium non-polar state and also exhibits a built-in electric field. This produces a shift in the O 2$p$ bands between unit cells in the STO, with the unit cell at the positively charged $LaO$-$TiO_2$ interface at the most negative potential relative to the Fermi level. The Cr 3$d$ and O 2$p$ bands in LCO also shift as the distance from the $LaO$-$TiO_2$ interface increases, with the bands moving towards the Fermi level. This type of behavior could in principle produce a strong enough electric field that would induce charge transfer and create a 2-DEG.[22] However, three unit cells of LCO appears to be below the critical thickness to produce this charge transfer, given that the Cr 3$d$ valence band lies below the conduction band at the $LaO$-$TiO_2$ interface.

Our calculations show that the oxygen octahedra in both the STO and LCO layers distort significantly, which can be attributed to the repeating positively and negatively charged interfaces in the superlattices. In the case of STO, the Ti ions are displaced away from the positively charged interface (LaO-TiO$_2$ interface), with asymmetric Ti-O apical bond lengths in the first five STO unit cells. The predicted difference between the long and short apical bond lengths is ~0.11 Å for these five TiO$_2$ layers. The interfacial STO unit cell does not exhibit this polar distortion but instead undergoes more pronounced octahedral tilting. This correlates with the flat band potential seen near unit cell 5 in **Figure 1**(e). Within the LCO layer, the Cr ions also distort in response to the built-in electric field. Given that the electric field points in the opposite direction in the LCO layer, the Cr-O bonds are distorted in the opposite direction. Again, the apical bonds closer to the positively charged interface are longer than those on the opposite side of the unit cell. Collectively, the built-in potentials across the superlattice should be detectable via x-ray photoelectron spectroscopy. Likewise, the predicted distortions to the octahedra should be observable using x-ray absorption spectroscopy and cation displacements should be seen in electron microscopy.

## 2.2. X-ray Spectroscopy

In agreement with *ab initio* modeling, core-level and valence band photoemission spectra measured *in situ* provide evidence for built-in potentials, as seen in **Figure 2**(a-e). Reference spectra from a single crystal STO substrate and a thick LCO film, each of which is in a nearly flat-band state, are also shown to illustrate the intrinsic peak widths of the core peaks. To mitigate the effects of photoelectron induced charging, an electron flood gun was used to neutralize the surface during measurement. This makes the binding energy scale inaccurate in an absolute sense, but we are able to correct for these effects by aligning all core level peaks such that the O 1*s* peak is at 530.3 eV, which is the measured value in doped *n*-type STO.[23]

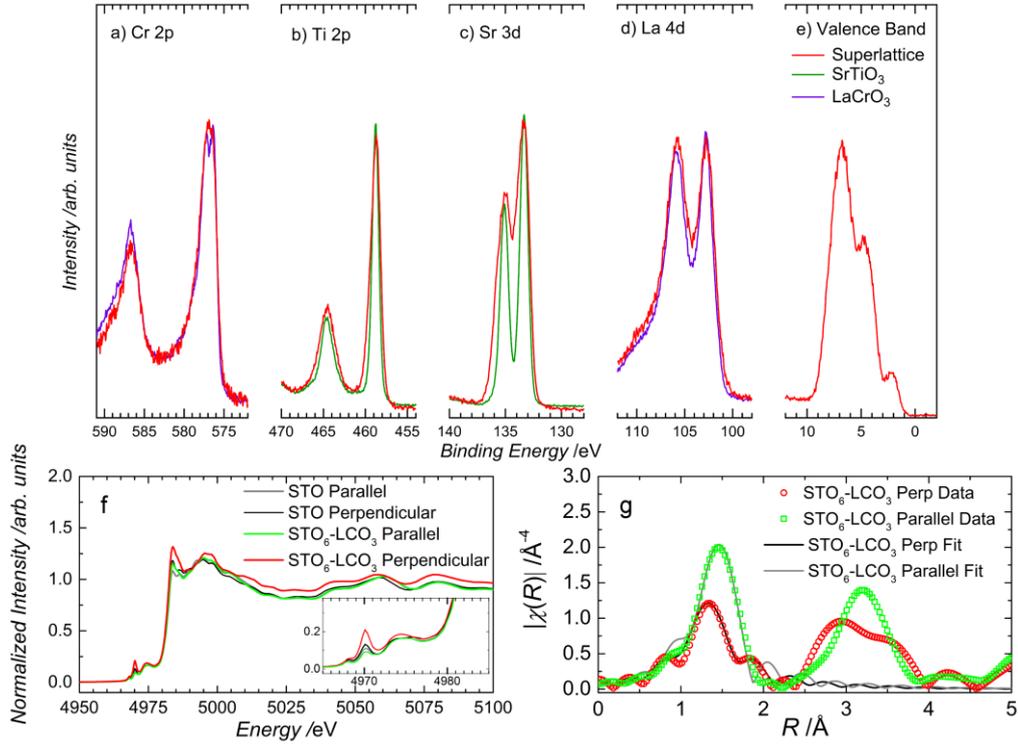

**Figure 2.** (a-e) In situ XPS spectra for a STO6-LCO3 superlattice and relevant SrTiO3 and LaCrO3 reference spectra. (f-g) XANES data for superlattices. f) Ti K edge spectra for a STO6-LCO3 superlattice and a reference STO film on LSAT; g) Magnitude of the Fourier transforms for the EXAFS obtained for the STO6-LCO3 sample with fits to the Ti-O contribution.

Each superlattice core peak shows significant broadening relative to the associated reference spectrum, a result attributable to variations in binding energy with depth relative to the Fermi level.[19,20] To estimate the potential drop across the layers of the superlattice, we model the peak broadening using the flat-band reference spectra for each layer.[19] This approach has been applied previously to examine the interfacial two-dimensional electron gas (2-DEG) between LAO and STO to determine if built-in potentials are present at the interface.[20,21,24] A notched band is expected on the STO side of the interface to confine electrons within the 2-DEG and a built-in potential should be present throughout the LAO layers if an electronic reconstruction occurs. However, neither of these has been observed. In contrast, a built-in potential gradient has been measured in LCO films at LCO-STO (001) interfaces.[19] These results are shown in Figure S2 and are discussed further in the supplemental information. We estimate a potential drop of roughly 0.9(2) V across both the STO and LCO layers of the superlattice near the film surface. Such a potential drop is in reasonable agreement with the theoretical prediction of 1.5 eV in Figure 1(c). The vacuum termination of the film surface may make the surface potential slightly different from that within the bulk of the superlattice, but using a laboratory source we are relatively insensitive to the potential gradient within buried layers. Further experiments using synchrotron hard x-ray photoelectron spectroscopy are planned to measure this effect deeper within the superlattice rather than at the surface and will be the subject of a future work.

Although the Cr 2$p$ and Ti 2$p$ superlattice peaks are broader than those from the reference specimens, they show no evidence of a change in oxidation states from the expected $Cr^{3+}$ and $Ti^{4+}$. $Ti^{3+}$ would produce a 2$p_{3/2}$ peak at a binding energy between 456 and 457 eV, as has been reported for La-doped $SrTiO_3$ with a sensitivity to ~2% $Ti^{3+}$ concentrations.[25] Likewise, the Cr 2$p$ peak does not show any clear shift in binding energy away from $Cr^{3+}$ that would be consistent with large concentrations (greater than ~10%) of $Cr^{4+}$.[26] These measurements suggest that there is no significant charge transfer occurring within the superlattice, but we are likely insensitive to small concentrations of oxygen vacancies that could occur at the interfaces. The presence of vacancies has been attributed to electronic Ti 3$d$ orbital polarization in STO-LAO superlattices, but the polarization effects are localized to the interface in that case.[15]

The valence band spectrum shows features that are consistent with the theoretical predictions (Figure 1(e)), with a Cr $3d$ $t_{2g}$ derived band near the Fermi level and the O $2p$ derived band from both the STO and LCO layers at higher binding energies. There is no apparent density of states at 0 eV, which is the nominal Fermi level after aligning the O $1s$ peak to 530.3 eV. This does not preclude the presence of small concentrations of free carriers in the film, but suggests that there is no large scale charge transfer to produce an interfacial 2DEG as has been seen in other XPS valence band measurements.[27] Based on the potential drops modeled from the core-level spectra, we constructed a simulation of the heterojunction valence band using reference STO and LCO spectra. The model (Figure S3) shows excellent agreement with the experimental results, accurately predicting the valence band broadening relative to the reference spectra. Taken in aggregate, the XPS analysis provides strong evidence for the presence of built-in electric fields within both layers in the superlattice.

To examine the local bonding environment of specific elements, polarization-dependent x-ray absorption measurements of the transition metal *K* edges were performed on the three superlattices, a reference STO film grown on LSAT, and a polycrystalline LCO film grown on $SiO_2$. The key results for the x-ray absorption near-edge spectroscopy (XANES) and extended x-ray absorption fine structure (EXAFS) are shown in **Figure 2**(f-g) for the $STO_6$-$LCO_3$ sample (see Figure S4(a) for other samples). The Cr *K* edges for $STO_6$-$LCO_3$ are presented in Figure S4(b) and show no deviation from the $Cr^{3+}$ oxidation state in LCO. A strong enhancement of the pre-edge intensity is seen in the Ti *K* edge spectra at ~4970 eV for the superlattice samples in the perpendicular polarization for all three samples relative to the STO film grown on LSAT under the same epitaxial strain. In contrast, the pre-edge peak is unchanged in each of the samples for the in-plane polarization, with the superlattice samples showing features very similar to those of the STO reference film. The superlattice samples also show different features at ~4975 eV in perpendicular polarization not seen in either the parallel polarization or in the STO control.

The enhanced Ti *K* shell pre-edge peak at ~4970 eV for the superlattices with perpendicular polarization indicates cation displacement normal to the interface.[28] Based on classic molecular orbital theory, pre-edge features for perfect octahedral coordination at the *K* edge are forbidden due to dipole selection rules, because the transition from a 1*s* orbital to a 3*d* orbital has a change in total angular momentum, $\Delta J$, of +2.[29] However, pre-edge features are still observed in ideal $TiO_6$ octahedra in a variety of compounds due to quadrupole transitions and *p-d* hybridization.[30,31] The intensity of the pre-edge peaks in both polarizations for the STO reference sample grown on LSAT is consistent with what is seen for $CaTiO_3$ in ideal octahedral symmetry.[30] However, the enhancement of the pre-edge peak seen for the perpendicular polarization in the superlattices is commonly found in cases where inversion symmetry in the octahedron is broken, such as in ferroelectric $BaTiO_3$ and $PbTiO_3$.[32] A similar response has been observed in epitaxial STO films grown on Si, where a polarization has also been observed.[28] Jiang et al. [31] showed that asymmetric Ti-O bond lengths increase the amount of $p$-$d$ hybridization and enhance the pre-edge peak. This suggests that the Ti octahedra within the STO layers of the superlattices are distorted in a manner consistent with the theoretical predictions from the DFT model.

The EXAFS provides quantitative information about the Ti distortions. Fourier transforms (FT) of the data, $\chi(R)$ show a strong polarization dependence **Figure 2**(g). The first peak in $\chi(R)$ is due to nearest-neighbor Ti-O bonding. If the Ti is shifted normal to the film plane due to a polar distortion, then there will be two Ti-O distances contributing to this peak in perpendicular polarization whereas the data for in-plane polarization will reveal a single bond length. The interference between photoelectrons backscattered from O ligands at the two bond lengths in the perpendicular data results in a dramatic reduction of the first-shell peak intensity in the FT. A model for this was constructed using the FEFF code,[33] and first shell fits were made using with a single Ti-O distance for the parallel data and two Ti-O distances for the perpendicular data to determine the bond lengths. A simple two-distance model yields a good fit to the polarization dependence, as shown in the figure. For the perpendicular data, the best fit for the Ti-O bonds gave a splitting of 0.20(3) Å. Similar values were observed for the $STO_8$-$LCO_4$ superlattice and with a reduced value of 0.13(8) in the $STO_4$-$LCO_2$ superlattice (Table S1 of the supplemental information). The reduction in the $STO_4$-$LCO_2$ superlattice may be due to the greater impact of interfacial intermixing in such a thin LCO layer. Additionally, based on the DFT model, the difference in apical Ti-O bond lengths at the positively charged interface is reduced, so with a greater interfacial density the mean difference in bond length may be reduced. Given the large depth sensitivity of the Ti $K$ edge measurement in fluorescence mode, the values represent a measure of the polarization throughout the superlattice, rather than at a single interface, indicating that the polarization persists throughout the sample.

## 2.3. Electron Microscopy

We use STEM-HAADF to directly visualize the ionic displacements with high spatial resolution to corroborate our x-ray spectroscopy evidence for a polarization. **Figure 3**(a) shows a representative aberration-corrected STEM-HAADF cross-section of the sample, marked with the position of the *A*- and *B*-site columns in the STO buffer layer. Displacements were measured by averaging the HAADF profiles from 10-20 *A*- and *B*-site column positions; similar results are found in several parts of the film, as shown in the supplemental information. We and others have shown that, using this approach, it is possible to measure ferroelectric polarization in $PbZr_{1-x}Ti_xO_3$ (PZT) with picometer precision that is unmatched by other techniques.[34–36] Similar examinations of the LAO-STO interface using STEM have shown that the $TiO_6$ octahedra near the interface are distorted to produce an off-center Ti cation displacement.[37,38] This is a sign of local polarization, but the polarization decays over a few unit cells within the STO lattice in the case of the LAO-STO interface.[39] **Figure 3**(b) shows the average *A*- and *B*-site intensity profiles from the columns marked in **Figure 3**(a). The arrows indicate the long ($\delta_{IL}$) and short ($\delta_{IS}$) displacements of the *B*-site cations relative to the edges of each unit cell. For a centrosymmetric cell we expect that $\delta_{IL} \approx \delta_{IS}$, but this is clearly not the case, as shown in **Figure 3**(c). We define the Ti cation displacement from the center of each unit cell as $\delta_{Ti}=(\delta_{IL}-\delta_{IS})/2$, and find that it is non-zero throughout the STO layer.

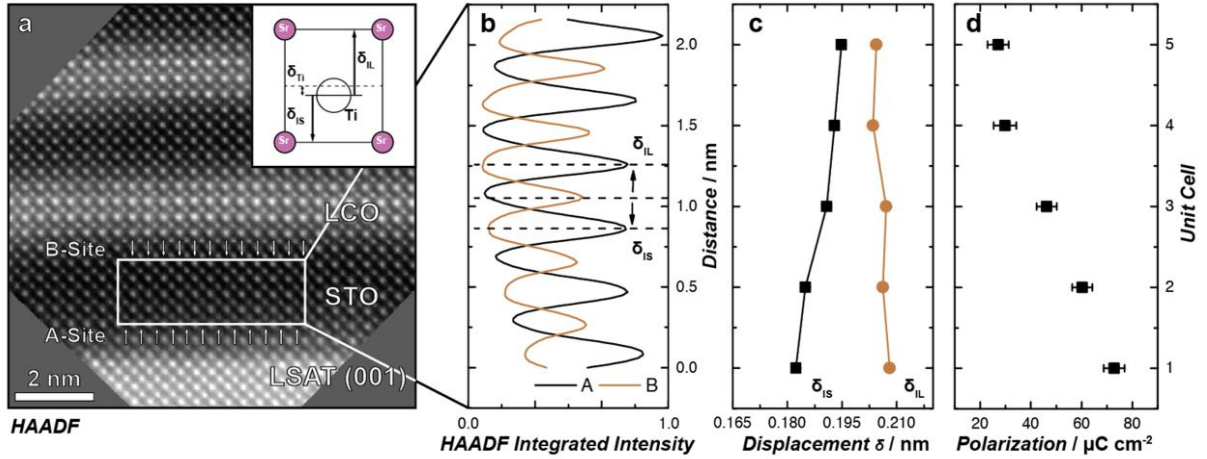

**Figure 3.** Measurement of local polarization. a) Drift-corrected representative STEM-HAADF micrograph of the STO buffer layer cross-section (45° scan direction); the arrows mark the A- and B-site cation columns; Inset) Geometry used to calculate the displacement vectors in b. b) Average intensity profiles of the *A*- and *B*-site columns in a. c) Measurement of the short and long displacement vectors for each unit cell. d) Estimate of local polarization for each unit cell. Error bars are calculated from the standard error of the Gaussian fits to each column position.

We consider these results in light of our electronic structure calculations and are able to directly relate the observed Ti displacements to the simulated ones. Doing so allows us to calculate the polarization from Born effective charge and atomic displacements within each layer of the crystal. We estimate the relationship between the effective out-of-plane polarization and Ti displacement as $P_S \approx \gamma \, \delta_{Ti}$, where $\gamma = 5663 \, \mu C \, cm^{-2} nm^{-1}$, as determined from DFT calculations (see supplemental for more details). This allows us to calculate the polarization across each unit cell, as shown in **Figure 3**(d). Our results reveal that the polarization is largest near the STO / LSAT interface, a value of 73(5) $\mu C \, cm^{-2}$. However, moving away from the interface toward the middle of the STO, the polarization drops to 46(5) $\mu C \, cm^{-2}$, finally reaching a minimum of 27(6) $\mu C \, cm^{-2}$ at the LCO / STO interface. This result is in agreement with our DFT calculations, which predict an average polarization of 32.5 $\mu C \, cm^{-2}$, and is greater than the room temperature polarization of $BaTiO_3$ (26 $\mu C \, cm^{-2}$).[40] Measurements in other regions of the superlattice show similar polarizations. Our results plainly show that the built-in potential can give rise to a polarization in these materials.

## 3. Discussion

The observed polarization throughout a superlattice consisting of non-ferroelectric perovskites is unusual and has not been previously observed. Such a polarization is not expected except in the case of alternating charged interfaces, as these produce electric fields in each STO and LCO layer of the superlattice. Polar distortions in STO have been observed near the interface in LAO-STO heterostructures, but they did not persist throughout the sample as they do here.[15,37,38] In the work of Park et al.[15], orbital polarization in STO was estimated to be screened over 1-2 unit cells away from the interface and was attributed to oxygen vacancies. Meanwhile, in a single LAO-STO heterojunction studied by STEM, the polarization in STO near the interface decayed to zero over roughly 2-3 unit cells in the works of Cantoni et al[37] and Jia et al.[38] The behavior in the LAO-STO system has been attributed in part to the role that interfacial oxygen vacancies play in mediating the polarization, though questions remain as to the physical mechanisms involved.[15,16] XPS depth profile analysis of our samples indicated that the stoichiometry of the STO layer throughout the sample was accurate to ~1-2%, suggesting that the polarization is not attributable to polar cation vacancies, as has been seen in other works.[41] A detailed study of the role of other interfacial defects on the potential gradient and induced polarization will be the subject of a future work.

A key difference between our STO-LCO superlattices and analogous STO-LAO samples is the nature of the band alignment between the two materials. In the case of STO-LAO, both materials are band insulators with the top of the valence band defined by O 2$p$ electronic states and small valence band offsets of a few tenths of an eV have been measured.[20,21] In the case of STO-LCO, however, LCO is a Mott insulator with a valence band offset 2.0-2.5 V relative to STO depending on LCO film thickness.[19] This may help to promote long range polarization and a built-in potential gradient within STO that is not observed in LAO/STO. The induced polarization in STO-LCO superlattices may be of particular interest for charge separation in these materials, as interfacial Cr-O-Ti bonds have been predicted to exhibit visible light optical transitions in superlattices.[18] Alloys of STO and LCO have demonstrated such behavior in both powder and thin film form,[42,43] supporting the theoretical predictions on the Cr-O-Ti bonds. The built-in electric field in these superlattices would be an excellent means of charge separation for photovoltaic or photocatalytic applications. Future studies exploring photoconductivity enhancement in these materials when compared to bulk alloys would help to elucidate these possibilities.

## 4. Conclusion

To summarize, our theoretical and experimental results reveal that alternating positively (LaO-TiO$_2$) and negatively (SrO-CrO$_2$) charged interfaces in LCO-STO superlattices induce a large polar distortion in the STO layer when the films are synthesized with asymmetric heterojunctions. *In situ* x-ray photoelectron spectroscopy measurements show core-level peak broadening consistent with a built-in potential difference of approximately 0.9 V across both the STO and LCO layers of the superlattice. These electric fields could be used to separate electron-hole pairs generated at interfacial Cr-O-Ti bonds, which exhibit visible light absorption.[18,42] Polarized x-ray absorption near edge spectroscopy at the Ti *K* edge are consistent with the displacement of Ti cations and the strong pre-edge feature is consistent with a polar distortion in the TiO$_6$ octahedra of STO. Aberration-corrected STEM-HAADF imaging confirms this polarization in the STO layers of the superlattice. By measuring the Ti cation displacements in the STO layer, we estimate that the polarization ranges from 27-73 $\mu$C cm$^{-2}$. Our results demonstrate that layering polar and non-polar materials can give rise to a strong polarization in the LCO / STO system comparable to that of ferroelectric BaTiO$_3$, and illustrate the extent to which the intrinsic properties of perovskites can be controlled and manipulated by artificial structuring.

## 5. Experimental Section

*Film Growth*: Superlattices were deposited by means of oxygen-assisted molecular beam epitaxy on $(LaAlO_3)_{0.3}(Sr_2AlTaO_6)_{0.7}$ (LSAT) substrates using metallic sources in differentially pumped effusion cells via a sequential shuttering technique.[44,45] The shuttered growth approach allows us to control the termination of each layer so that both $TiO_2^0$-$LaO^+$ and $CrO_2^-$-$SrO^0$ interfaces are present. The base pressure of the chamber was better than $5\times10^{-9}$ Torr and the films were grown in a molecular oxygen background pressure of $3\times10^{-6}$ Torr at 700 °C on LSAT substrates. The details of the flux calibration process are described in the Supplemental Information. A STO buffer layer a few unit cells in thickness was deposited on the LSAT to produce a $TiO_2$ termination for the film prior to the beginning of the superlattice growth.

*X-ray Photoelectron Spectroscopy:* After growth, the samples were transferred under ultra-high vacuum to an appended XPS analysis chamber. The system is equipped with a monochromatic Al K$\alpha$ x-ray source and VG/Scienta R3000 analyzer. Because of the insulating nature of the LSAT substrates, the superlattices were isolated from ground and an electron flood gun was required to neutralize the positive charge resulting from photoelectron ejection, thus making the absolute determination of binding energies impossible. To better estimate the true binding energies, all spectra were shifted by a constant value required to align the O 1$s$ peak to 530.3 eV, the value we measure for clean Nb-doped STO(001) in a flat band condition.

*X-ray Absorption Spectroscopy:* Measurements were performed on the Cr and Ti $K$ edges at the Advanced Photon Source on beamlines 20-BM and 20-ID, both using a Si (111) monochromator with energy resolution of about 0.8 eV. Energy calibration was done by setting a Cr metal standard edge to 5989.0 eV. Scans for both the in-plane (parallel) and out-of-plane (perpendicular) polarizations were taken. For both polarizations the samples were set at a small angle (5-7°) to the focused beam and spun around the sample normal to avoid interference from sample Bragg reflections. An STO film on LSAT was also measured using the same conditions to determine the effect of epitaxial strain on the pre-edge features. The data was analyzed using the Demeter XAS software suite.[46]

*Electron Microscopy:* The STEM-HAADF image in **Figure 3**(a) is the average of a relatively high-speed time series of 39 acquisitions acquired at ~0.1s intervals (0.4 $\mu$s per pixel) to improve the signal-to-noise ratio. The individual frames were processed using both rigid and non-rigid correction routines to correct both sample drift and scan noise.[47] STEM-HAADF images were collected on a $C_S$-corrected Nion UltraSTEM 100 operated at 100 keV, with a convergence angle of 30 mrad. The HAADF inner and outer collection angles were 82 and 190 mrad, respectively. The data set was first rigid registered to eliminate any sample or stage drift.[47] High-frequency scan-noise was then compensated using the Jitterbug software (HREM Research).[48] Importantly, the scan-noise was compensated in each individual frame of the series before averaging across the series. The data were not smoothed or filtered in any way. The STEM-HAADF images were processed as follows: stage drift offset vectors were determined by windowing each image in real-space followed by cross-correlation. These offset vectors were used to align un-windowed data before further robust row-locked non-rigid registration. Both tasks were performed using the Smart Align software.[47] In addition, we have conducted measurements with the scan direction both parallel to the interface and at 45° to account for possible scanning artifacts (see Figure S5 for additional scan, which shows similar results).

*Density Functional Theory:* The structure and properties of the ideal $[SrTiO_3]_6/[LaCrO_3]_3$ hetero-structures were examined using computational modeling and density functional theory (DFT). We employed the exchange-correlation functional by Perdew–Burke–Ernzerhof[49] and modified for solids (PBEsol)[50], plane wave basis set with the energy cutoff of 500 eV and the projected augmented waves method[51] implemented in the Vienna Ab initio Simulation Program (VASP).[52,53] The heterostructures were represented using the periodic model and the $\sqrt{2}a_0 \times \sqrt{2}a_0$ lateral cell, where $a_0$ corresponds to the lattice constant of a cubic perovskite. The lateral lattice parameters were constrained to 5.4702 A, which corresponds to the film being coherently strained to the LSAT substrate with the lattice constant of 3.868 A. The out-of-plane parameter *c* was varied so as to minimize the total energy of the idealized hetero-structure. The fractional coordinates were re-optimized at every value of c. These calculations were conducted for the $U = 3.0$ eV for 3$d$ states of both Cr and Ti and resulted in $c = 35.6$ Å, which is within 0.8% of the experimental value, and consistent with the normal overestimation of lattice parameters in DFT. The 4×4×1 Monkhorst-Pack $k$-grid centered at the Γ point was used in all calculations. First, the total energy of the idealized hetero-structure was minimized with respect to the lattice parameters and the fractional coordinates. The PBEsol+$U$ approach[50,54] was adopted unless stated otherwise; $U_{Ti} = 8.0$ eV and $U_{Cr} = 3.0$ eV, which reproduce the one electron band gaps in STO and LCO, respectively, were used.

**Supporting Information**
Supporting Information is available from the Wiley Online Library or from the author.

**Acknowledgements**
RBC was supported by the Linus Pauling Distinguished Post-doctoral Fellowship at Pacific Northwest National Laboratory (PNNL LDRD PN13100/2581). SRS, MEB and SAC were supported by the U.S. Department of Energy (DOE), Basic Energy Sciences (BES), Division of Materials Sciences and Engineering under Award #10122. PVS. and P-VO were supported by the LDRD Program at PNNL. PNNL work was performed in the Environmental Molecular Sciences Laboratory, a national science user facility sponsored by the DOE Office of Biological and Environmental Research. Sector 20 facilities at the Advanced Photon Source (APS), and research at these facilities, are supported by the US DOE BES, the Canadian Light Source and its funding partners, the University of Washington, and the APS. Use of the APS, an Office of Science User Facility operated for the DOE Office of Science by Argonne National Laboratory, was supported by the DOE under Contract No. DE-AC02-06CH11357. Electron microscopy was carried out in parts at the SuperSTEM Laboratory, the U.K. National Facility for Aberration-Corrected STEM, which is supported by the Engineering and Physical Sciences Research Council (EPSRC). The research leading to these results has received funding from the European Union Seventh Framework Programme under Grant Agreement 312483 - ESTEEM2 (Integrated Infrastructure Initiative–I3).

Received: ((will be filled in by the editorial staff))
Revised: ((will be filled in by the editorial staff))
Published online: ((will be filled in by the editorial staff))

**Through the tailored growth of alternating positively and negatively charged interfaces,** emergent polarization is observed in $SrTiO_3$-$LaCrO_3$ superlattices. A built-in electric field is predicted via density functional theory and measured experimentally via x-ray photoelectron spectroscopy. Measurements of Ti cation displacements within oxygen octahedra via x-ray absorption spectroscopy and electron microscopy demonstrate a polarization comparable to that of ferroelectric $BaTiO_3$.

**Keyword: Complex oxide superlattices, x-ray photoelectron spectroscopy, x-ray absorption spectroscopy, scanning transmission electron microscopy, density functional theory**

R. B. Comes,1 S. R. Spurgeon,1 S. M. Heald,2 D. M. Kepaptsoglou,3 L. Jones,4 P. -V. Ong,1 M. E. Bowden,5 Q. M. Ramasse,3 P. V. Sushko,1 and Scott A. Chambers1*

**Interface-induced Polarization in $SrTiO_3$-$LaCrO_3$ Superlattices**

ToC figure

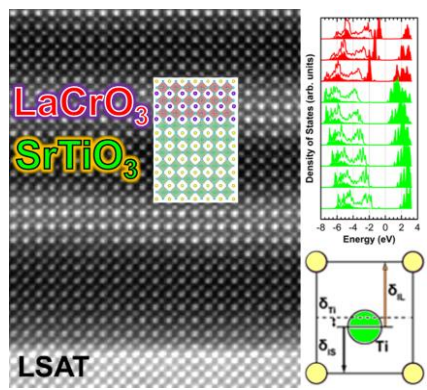

# Supporting Information

**Interface-induced Polarization in SrTiO$_3$-LaCrO$_3$ Superlattices**

*Ryan B. Comes,[a]\* Steven R. Spurgeon,[a] Steve M. Heald,[b] Despoina M. Kepaptsoglou,[c] Lewys Jones,[d] Phuong-Vu Ong,[a] Mark E. Bowden,[e] Quentin M. Ramasse,[c] Peter V. Sushko,[a] and Scott A. Chambers[a]\**

[a]Fundamental and Computational Sciences Directorate, Pacific Northwest National Laboratory, Richland, Washington 99352, United States

[b]Advanced Photon Source, Argonne National Laboratory, Argonne, Illinois 60439, United States

[c]SuperSTEM, SciTech Daresbury Campus, Daresbury, WA44AD, United Kingdom

[d]Department of Materials, University of Oxford, Oxford OX13PH, UK

[e]Environmental Molecular Sciences Laboratory, Pacific Northwest National Laboratory, Richland, Washington 99352, United States

*Film Growth and Calibration*

The initial source calibration was performed using a quartz crystal rate monitor. Prior to superlattice growth, a homoepitaxial STO film was grown on an STO substrate to calibrate the sources to within ~1-2% based on RHEED oscillations.[S1,S2] After completing the STO film, an LCO film was immediately deposited on the same sample through co-deposition from the La and Cr sources. The RHEED pattern and reconstruction were monitored to calibrate the flux to better than 5% precision.[S3] The LSAT substrate was then loaded into the chamber with the sources at growth temperature and the superlattice was grown using a sequential shuttering process for all four sources.

During growth, the film was grown via a shuttered growth process following the sequence of: Sr-Ti-…Sr-Ti-La-Cr…La-Cr-Sr-Ti… … with one shutter opening corresponding to one monolayer AO or BO$_2$ plane in the superlattice. Each shutter was held open for 43 seconds, corresponding to a growth rate of ~2.7 Å/min. The RHEED pattern after growth is shown in **Figure S1**, with sharp streaks indicating layer-by-layer growth throughout the process.

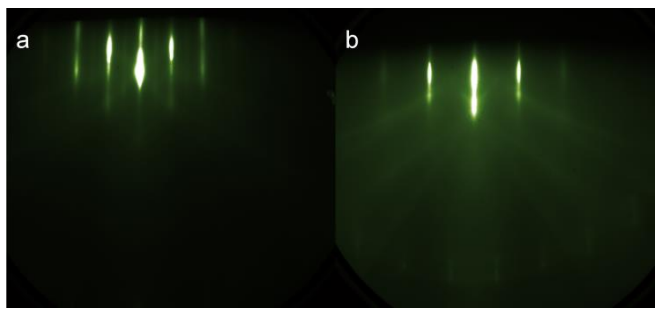

**Figure S1.** Post-growth RHEED pattern along (a) [10] azimuth and (b) [11] azimuth. Reciprocal space map measurements of the structural properties of the STO$_6$-LCO$_3$ sample are shown below in Figure S2. The film is coherently strained to the LSAT substrate, with an in-plane lattice parameter of 3.868 Å. Because STO and LCO have different bulk lattice parameters, discerning the lattice parameter of the individual layers is challenging. To accomplish this, we have conducted geometric phase analysis (GPA) to measure lattice parameter changes parallel to the film growth direction. Figure S3(a) shows a representative STEM-HAADF micrograph of the film structure. From this image we choose a bulk LSAT reference region to use as our baseline for strain measurements. We then calculate local changes in the Fourier components of lattice fringes to estimate the local strain state relative to the LSAT reference with nanometer precision.[S4,S5] Figure S3 shows the calculated in-plane strain component ($\varepsilon_{xx}$), which exhibits a uniform intensity throughout, indicating that the film is coherently strained in plane. Figure S3(c-d) show the calculated out-of-plane strain component ($\varepsilon_{xx}$) and corresponding strain profile averaged in the plane of the film. We observe a regular modulation with the same periodicity as the HAADF image, indicating changes in out-of-plane lattice parameter; we find a tensile strain of ~1.5% relative to the bulk LSAT lattice parameter, indicating that $c \approx (3.868 \text{ Å})(1.015) \approx 3.92$ Å, in agreement with our XRD measurements and DFT calculations.

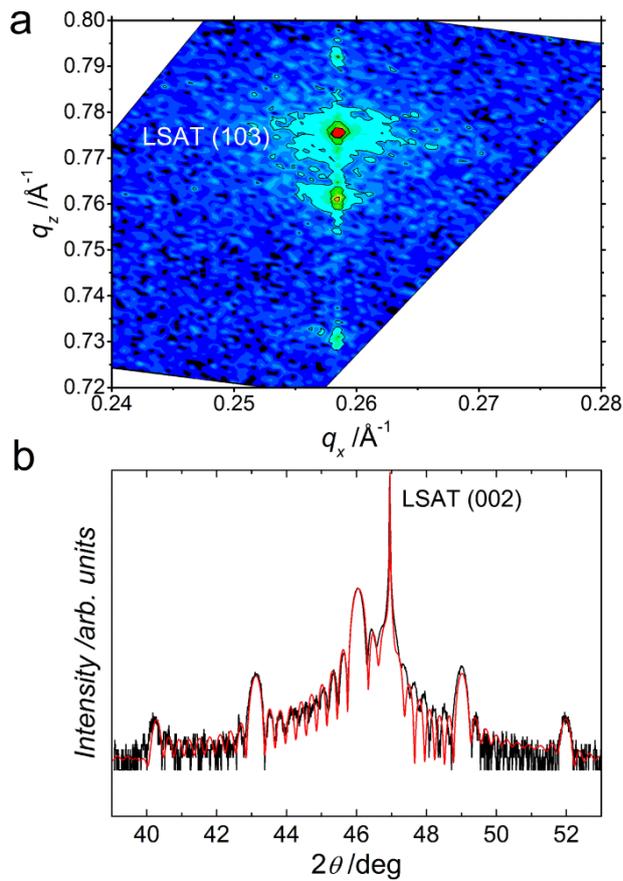

**Figure S2**. a) Reciprocal space map of STO6-LCO3 superlattice about LSAT (103) peak; b) Fit (red) to out-of-plane data around LSAT (002) peak.

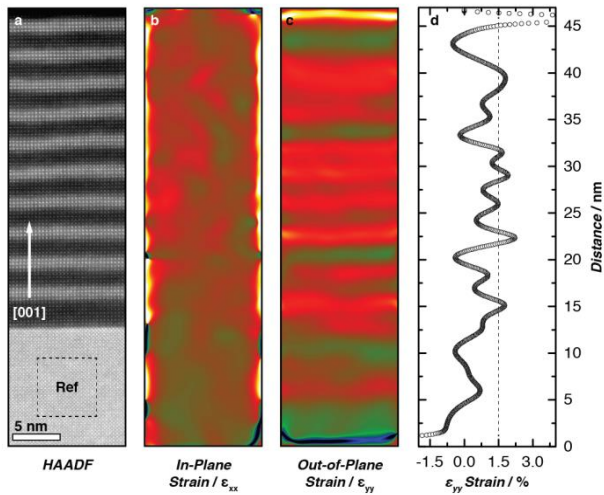

**Figure S3**. Geometric phase analysis of STO6-LCO3 superlattice. a) Representative STEM-HAADF image of the heterostructure, overlaid with the reference region used for strain analysis. b) In-plane strain component ($\varepsilon_{xx}$), showing that the film is coherently strained. The bright regions around the edges are an artifact of the measurement. c) Out-of-plane (normal to film surface) strain component ($\varepsilon_{yy}$), showing a an abrupt modulation in the $c$ axis lattice parameter commensurate with the film layers. d) Line profile integrated in the plane of the film, revealing an out-of-plane expansion on the order of ~1.5% tensile strain relative to bulk

LSAT. The noise in this profile is likely the result of residual scanning artifacts during the STEM measurement.

*Density Functional Theory Model*

To estimate any internal electric field in the system, we defined a regular grid of ~400 points along the c-axis of the hetero-structure and calculated the average electrostatic potential in the a–b plane for every such grid point. This potential was then averaged over the length of one LCO unit cell period to produce the potential shown in Figure 1(c) in the main text. This averaging artificially produces small oscillations within the STO region of the superlattice due to the differing lattice parameters between STO and LCO. If the STO lattice parameter is used instead, the potential does not oscillate in STO but does within LCO. This is shown below in Figure S4.

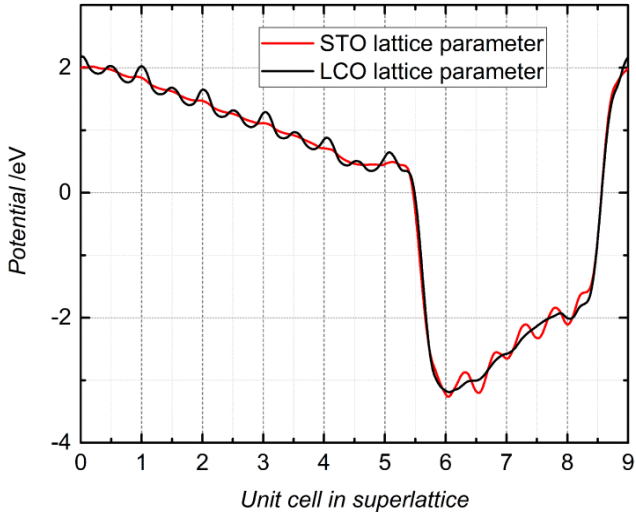

**Figure S4.** Potential maps with different averaging windows. Window over $SrTiO_3$ unit cell is red and $LaCrO_3$ unit cell is black.

To investigate the layer-resolved polarization in the superlattice, we compute the local polarization, $P$, in each unit-cell-thick layer, defined as:

$$P_{\alpha,local} = \frac{1}{\Omega}\left\{\sum_{Ti(Cr)} Z^*_{\alpha,Ti(Cr)} u_{\alpha,Ti(Cr)} + \sum_{O_{eq}} Z^*_{\alpha,O_{eq}} u_{\alpha,O_{eq}} + \frac{1}{2}\sum_{Sr(La)_+,Sr(La)_-} \left(Z^*_{\alpha,Sr(La)_+} u_{\alpha,Sr(La)_+} + Z^*_{\alpha,Sr(La)_-} u_{\alpha,Sr(La)_-}\right) + \frac{1}{2}\sum_{O_{ap+},O_{ap-}} \left(Z^*_{\alpha,O_{ap+}} u_{\alpha,O_{ap+}} + Z^*_{\alpha,O_{ap-}} u_{\alpha,O_{ap-}}\right)\right\}$$
**(S1)**

where $\Omega$ is the volume of a $\sqrt{2}\times\sqrt{2}\times1$ unit-cell, $\alpha$_is a Cartesian direction $(x, y, z)$, $Z^*_{\alpha,i}$ is the Born effective charge of atom $i$ along $\_\alpha$, $u_{\alpha,i}$ is atomic displacement of atom is with respect to the centrosymmetric cell defined by the corner Sr(La). $O_{eq}$ and $O_{ap}$ represent O atoms at the equatorial and apical positions of an octahedron, respectively. $O_{ap+}$ and $O_{ap-}$ are the $Oap$ atoms at top and bottom of the unit-cell, repectively. Similarly, $Sr(La)_+$ and $Sr(La)_-$ represent the Sr(La) atoms at top and bottom of the unit-cell. The Born effective charges are approximated by those in bulk $SrTiO_3$ for STO layers and those in bulk $LaCrO3$ in the LCO layers.

*X-ray Photoelectron Spectroscopy Analysis*

The magnitude of the built-in electric field within each unit cell of the superlattice may be estimated using models of the core level peaks in flat band conditions. This is accomplished through the use of an STO substrate reference and LCO thick film measured under the same conditions. In the case of a built-in field, the core level peaks should broaden due to differing binding energies in each unit cell of the superlattice. Since the intensity of photoexcited electron emission is exponentially dependent on the depth below the surface, we can model each unit cell independently by summing the reference spectra together after applying energy shifts to account for the built-in field. This model can be represented mathematically as:

$$I(E) = \sum_{j=0}^{n-1} I_0(E - j\Delta E) \exp\left(-\frac{jc}{\lambda}\right) \quad \text{(S2)}$$

where $I(E)$ is the measured intensity, $I_0(E)$ is the intensity of the reference spectrum, $n$ is the number of unit cells in the model, $\Delta E$ is the potential drop per unit cell, $c$ is the out-of-plane lattice parameter, $j$ is the unit cell a distance $jc$ from the film surface, and $\lambda$ is the inelastic mean free path of the photoexcited electron. In practice, this is challenging, as various assumptions must be made regarding cation intermixing, the inelastic mean free path, and the fact that the energy varies linearly with depth. We assume a mean free path, $\lambda$, of 15 Å throughout the film, though this somewhat simplistic given the differing band gaps and valence electron densities in LCO and STO. For both layers we model the 3 unit cells closest to the film surface, the 3 u.c. STO capping layer and the final 3 u.c. LCO layer. The more buried layers will be strongly attenuated and have a significantly smaller contribution to the spectra. Models for the Sr 3*d*, Ti 2*p*, La 4*d*, and Cr 2*p* core level peaks are shown below in Figure S5(a-d). Various values for *ΔE* are modeled and the best fit to the data is estimated. For both the Sr 3d and Ti 2p core level spectra, a potential drop of 0.5 eV/u.c. The La 4d level for the LCO layer shows good agreement with a drop of 0.4 eV/u.c. In the case of the Cr 2*p* peak, the inherent width due to multiplet splitting increases the uncertainty in the analysis, but the model agrees well with the 0.4 eV/u.c. measured for La 4*d*. These values are consistent with what has been observed previously in LCO films grown on STO.

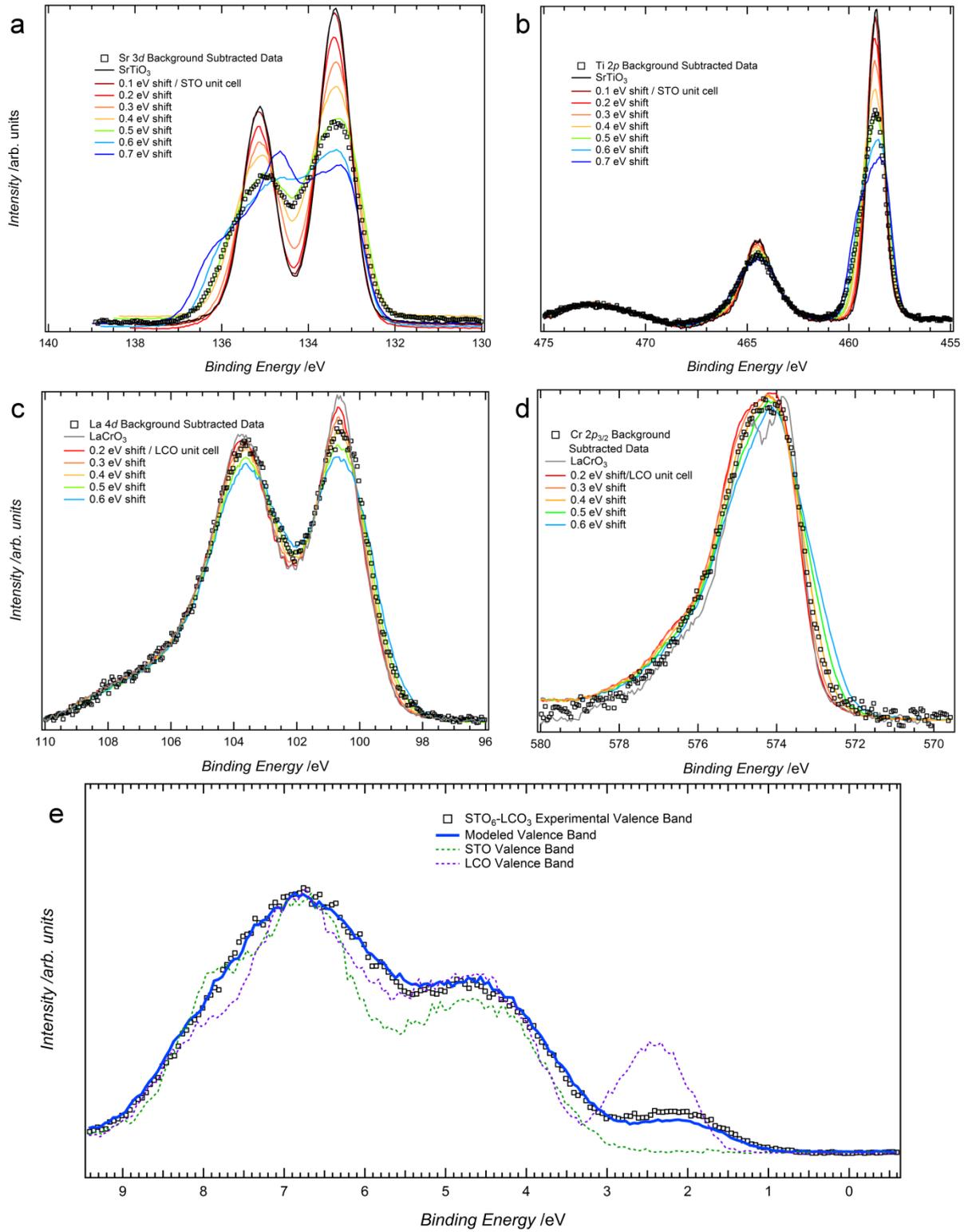

**Figure S5.** Core level peak broadening models with various potential drops per unit cell for the (a) Sr 3*d*, (b) Ti 2*p*, (c) La 4*d* peaks, and (d) Cr 2*p*. (e) Model of valence band spectrum assuming potential drops measured in core levels.

Given the estimated potential drops from the core level spectra, we can model the valence band spectrum by summing LCO and STO reference valence band spectra over top most six unit cells using the same methodology. The band alignment was chosen to match that of the density functional theory model in Figure 1(c). These results are shown in Figure S5(e), along with the measured valence band and the LCO and STO references. The measured superlattice valence band is aligned by placing the O 1$s$ peak at 530.3 eV,[S12] while the remaining spectra are aligned so that the peaks at 7 eV match. The model shows excellent agreement with the valence band spectrum and accurately replicates the broadened Cr 3$d$ peak between 3 eV and 1 eV. These results are highly promising, but are somewhat idealized given that they are not necessarily a unique solution to the potential in the system. Ongoing measurements using synchrotron standing-wave XPS measurements are focused on providing more detailed analysis and will be the subject of a future work.

*X-ray Absorption Spectroscopy*

The Ti K edge XANES data for both all three superlattice samples is shown below in Figure S6(a). All three superlattice samples exhibit similar changes in the pre-edge structure in the perpendicular polarization, without corresponding changes to the in-plane polarization. This suggests that the ferroelectric distortion occurs within the bulk of layers and not just near the interfaces. Were the polarization limited to the interfaces, we would expect the pre-edge peak intensity to scale inversely with the real space period of the superlattice. That is, the greater the density of interfaces the larger the pre-edge enhancement. Instead, we observe that the pre-edge intensity is slightly smaller for the $STO_4$-$LCO_2$ superlattice and is essentially unchanged for the $STO_6$-$LCO_3$ and $STO_8$-$LCO_4$ superlattice.

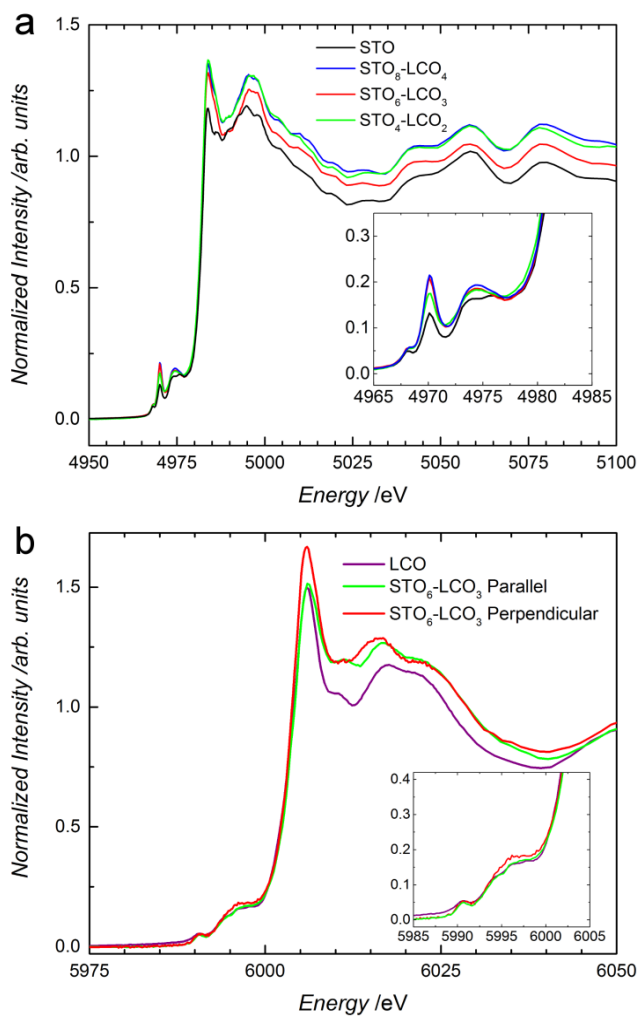

**Figure S6**. a) Ti K edge XANES spectra for perpendicular polarization in three superlattice samples. b) Cr K edge XANES spectra for $STO_6$-$LCO_3$ in both polarizations with LCO reference.

Fits to measure the Ti-O bond length via extended x-ray absorption fine structure (EXAFS) measurements at the Ti K edge were performed using the Artemis software package and the FEFF code.[S11] The presence of the La $L_3$ edge at 5483 eV limits the Ti EXAFS data range which is available, but fits to the Ti-O bond length were possible. Because the data range was limited, the overall amplitude factor and $E_0$ value were first determined from a fit to the STO reference film measured at the same time. with a k-space window for the transform of 2-10 Å$^{-1}$. The fitting range in R was 1-2 Å. The resulting difference in the perpendicular bond length for the samples is a measure of the cation displacement in the system and directly proportional to the polarization. EXAFS fits to the difference in Ti-O bond length for all three samples are shown in **Table S1**. The differences in axial bond length, $\Delta R$, are slightly larger than the predicted values of 0.11 Å from the DFT models.

**Table S1**: The numerical results from fitting the first shell EXAFS.

| Sample | $R_{plane}$ | $\sigma^2$ | $R_{short}$ | $\sigma^2$ | $R_{long}$ | $\sigma^2$ | $\Delta R$ |
|---|---|---|---|---|---|---|---|
| STO$_4$-LCO$_2$ | 1.94(2) | 0.0003(10) | 1.89(4) | 0.002(4) | 2.02(4) | 0.006(4) | 0.13(8) |
| STO$_6$-LCO$_3$ | 1.94(2) | 0.0009(10) | 1.87(2) | -0.0009(30) | 2.07(3) | 0.004(6) | 0.20(3) |
| STO$_8$-LCO$_4$ | 1.95(2) | 0.000(1) | 1.85(2) | 0.0001(17) | 2.04(3) | 0.004(3) | 0.19(3) |

An LCO film previously grown on SiO$_2$ was also used as a reference for the Cr *K* edge.[S13] The Cr K edge data shown in Figure S6(b) indicate that there is no deviation from the Cr$^{3+}$ oxidation state when compared to the LCO reference. The rising edge and white line peak positions are essentially identical between the reference and the superlattice. The increased intensity of the white-line peak at ~6006 eV can be attributed to the confined nature of the LCO layer in the superlattice along the growth direction. An enhancement of the white-line intensity has been reported in layered compounds such as TaS$_2$ and WSe$_2$ previously, and was attributed to the narrow thickness of the layers and the anisotropy in the electronic states.[S14] There is a slight change in the pre-edge feature for the perpendicular polarization at ~5995 eV. Similar deviations are observed in the 8-4 and 4-2 superlattices, suggesting that the variation is statistically relevant. This slight distortion in pre-edge structure may be attributable to the predicted asymmetric bond lengths along the perpendicular direction, though theoretical and experimental studies on this are limited. The pre-edge distortion is qualitatively similar to what is seen in Cr$_2$O$_3$, which has distortions in the Cr-O bond length due to the corundum structure of the material.[S12] Thus, it seems plausible that distortions are present in the CrO$_6$ octahedra, though further corroboration through other means is needed to verify this.

*Ferroelectric Polarization Measurement*

As described in the main text, we have conducted STEM-HAADF imaging with the beam scanning direction both parallel and at 45º to the film interface to account for possible scanning distortions. The former case is shown in **Figure S7** and agrees well with the results presented in the main text. This image is the average of a relatively high-speed time series of 30 acquisitions acquired at ~1.5s intervals (2.5 μs per pixel); the individual frames were processed using both rigid and non-rigid correction routines to correct both sample drift and scan noise, as is described in the methods section of the main text.[S15] We observe a similarly large polarization on the order of $77 \pm 5$ μC cm$^{-2}$ at the STO / LSAT interface; this decays to $57 \pm 6$ μC cm$^{-2}$ in the middle of the STO, finally reaching a minimum of $50 \pm 7$ μC cm$^{-2}$ at the LCO / STO interface. This again suggests that interfacial strain may act to enhance the polarization at the film-substrate interface, resulting in a non-uniform distribution of polarization.

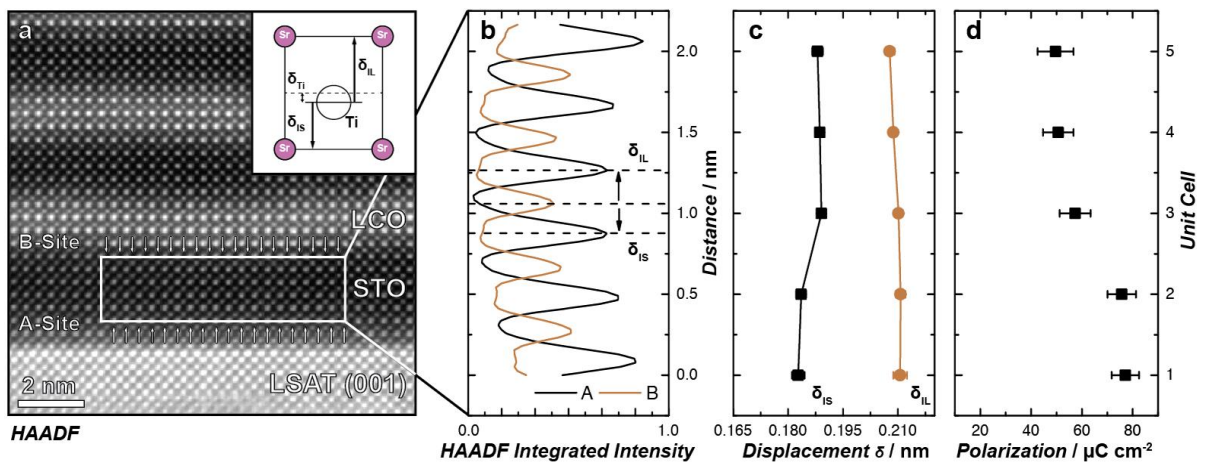

**Figure S7**. Measurement of local ferroelectric polarization. a) Drift-corrected representative STEM-HAADF micrograph of the STO buffer layer cross-section (scan parallel to interface); the arrows mark the *A*- and *B*-site cation columns. The inset indicates the geometry used to calculate the displacement vectors in b. b) Average intensity profiles of the *A*- and *B*-site columns in a. The long ($\delta_{IL}$) and short ($\delta_{IS}$) displacements of the *B*-site relative to the unit cell center are indicated by the arrows. c) Measurement of the short and long displacement vectors for each unit cell. d) Estimate of local polarization for each unit cell using the constant determined from DFT calculations. There is a clear decrease in polarization moving from the LSAT to LCO layers. Error bars are calculated from the standard error of the Gaussian fits to the atomic columns.